# Wideband Infrared Spectrometer for Characterization of Transiting Exoplanets with Space Telescopes

Keigo Enya

Institute of Space and Astronautical Science, Japan Aerospace Exploration Agency
3-1-1 Chuou-ku, Yoshinodai, Sagamihara, Kanagawa 252-5210, JAPAN

enya@ir.isas.jaxa.jp

Abstract

This paper presents a conceptual design for a spectrometer designed specifically for characterizing transiting exoplanets with space-borne infrared telescopes. The design adopting cross-dispersion is intended to be simple, compact, highly stable, and has capability of simultaneous coverage over a wide wavelength region with high throughput. Typical wavelength coverage and spectral resolving power is 1–13 micron with a spectral resolving power of ~ a few hundred, respectively. The baseline design consists of two detectors, two prisms with a dichroic coating and microstructured grating surfaces, and three mirrors. Moving parts are not adopted. The effect of defocusing is evaluated for the case of a simple shift of the detector, and anisotropic defocusing to maintain the spectral resolving power. Variations in the design and its application to planned missions are also discussed.

## 1. Introduction

The detailed study of exoplanets is an important issue in space science to understand how planetary systems were born, how they evolve, and, ultimately, to find biological signatures on these planets. Recently, monitoring the spectra (or spectral energy distributions) of transiting exoplanets, together with their central stars which are not spatially resolved, was pioneered, and this has proven to be a valuable method for revealing the atmospheric features of exoplanets. There are many absorption features of exoplanet atmospheres in the infrared wavelength region, e.g., $H_2O$ (1.13, 1.38, 1.9, 2.69, 6.2micron), $CO_2$ (1.21, 1.57, 1.6, 2.03, 4.25micron), $O_3$ (4.7, 9.1, 9.6micron), $O_2$ (1.27micron), $CH_4$ (1.65, 2.2, 2.31, 2.37, 3.3, 6.5, 7.7micron), $NH_3$ (1.5, 2, 2.25,

2.9, 3.0, 6.1, 10.5micron), and so on (e.g., Tinetti et al. 2012, Swain et al. 2010, and their references). For the characterization of transiting exoplanets by spectroscopic monitoring, stability of the observation is essential because the spectral features of exoplanets are derived as the difference between the transit−in and the transit-out. Therefore, next generation space−borne infrared telescopes, for instance, the Space Infrared telescope for Cosmology and Astrophysics (SPICA; e.g., Nakagawa 2010 and its references) and the James Webb Space Telescope (JWST; e.g., Clampin 2008), will be important platforms for such observations because of their large aperture, wide infrared wavelength coverage, capability of observations free from air turbulence, and high stability of optics at cryogenic temperature (Enya et al. 2007; 2012). With taking these features into consideration, this paper presents a conceptual design for a wideband infrared spectrometer designed specifically for observing transiting exoplanets with space−borne infrared telescopes.

## 2. Optical design

### 2.1. Baseline design

The spectrometer has to be both simple and compact, be highly stable, have a high throughput, and simultaneous coverage over a wide wavelength region. An example of a design that can achieve these features, Design #1, is presented in Fig.1. In order to study the atmospheres of explanets, e.g., composition, vertical distribution of physical properties, variability, and so on, the targets for wavelength coverage and spectral resolving power are set to be ~1−13micron and a few hundred, respectively. For the high stability, no moving parts are included in the optics. To realize simultaneous wideband coverage, the spectrometer has two channels: a short-channel (S-CH) and a long-channel (L-CH) to cover wavelengths shorter and longer than 5 microns, respectively. Furthermore, cross dispersion is adopted for the spectroscopy in each channel (Enya et al. 2011). A small number of optical devices in each channel makes the design simple and compact and promotes throughput. Table 1 shows a summary of this design.

For the optical design and analysis given in this paper, the optics start with an ideal point source on the focal plane, i.e., telescope aberration is not included for ease of understanding. The beam is collimated by an off-axis parabolic mirror, Mirror-C. A dichroic multi-layer coating is applied on the incident surface of a ZnS prism, Prism-L, and then the beam is split into the S-CH and L-CH. The incident angle at the surface of Prism-L is ~60 degree. The L-CH beam passes through Prism-L, while the S-CH beam enters a ZnSe prism, Prism-S. The dispersion powers of the prisms are used only to separate the orders of the spectrum in each channel. The spectral dispersion in each order is realized by grating patterns manufactured on the exit surfaces of the prisms. The dispersion angle of the prism and the grating are perpendicular to each other. Free-form optimized off-axis mirrors, Mirror-S and Mirror-L, are used for focusing S-CH and L-CH, respectively, onto detector arrays.

To simulate the spectral dispersion at the detectors, the telescope aperture is assumed to be circular with a diameter of 3m. Aberration caused by the telescope is not considered. The supposed pixel sizes of the detectors for S-CH and L-CH are 25micron × 25micron and 30micron × 30micron, respectively. The expected spectral dispersions at the detectors are shown in Fig.2, and spot diagrams are shown in Fig.3. The spectral resolving poer, $R$, and the Strehl ratio are summarized in Table A1 in the appendix. The available $R$ is limited to the minimum of the three values shown in the table; 1.22 × lambda / D (diffraction limit), 2 × pixel size (sampling limit), and RMS SPOT (aberration limit). Table A1 shows that $R$ for the L-CH is diffraction-limited for all wavelengths. $R$ for the S-CH is sampling-limited (i.e., under-sampling) at wavelengths less than ~1.8 micron and diffraction-limited at wavelengths greater than ~1.8micron. In actual use, it is expected that the issue of under-sampling is relaxed by telescope aberration and/or the effect of defocusing described in the following section.

## 2.2. Defocusing

For monitoring transiting exoplanets, defocusing the optics is often used to avoid flux saturation of the detectors and to make the observation robust over fluctuations in telescope pointing. To evaluate the effect of defocusing, the detectors were shifted in the simulation toward the focusing mirrors to decrease the Strehl ratio at the shortest wavelength in each channel to 0.2. As a result, the detectors were shifted by 4.00mm and 3.55mm in S-CH and L-CH, respectively. Spot diagrams of are shown in Fig.4. $R$ and Strehl ratio are summarized in Table A2 in the appendix. In this case, $R$ is limited by RMS SPOT at almost all wavelengths in both channels. As a result of defocusing, $R$ tends to be reduced and to become relatively uniform over the wavelength coverage of each channel. The under-sampling in S-CH found above is removed. In L-CH, the effect of defocusing on the Strehl ratio is more evident at shorter wavelengths. For example, the Strehl ratio reduces to ~0.2 at 5 microns, while it is still greater than 0.8 at 13 microns. This trend is preferable for the observation of transiting exoplanets because observed flux is dominated by the central star, and the saturation by the spectral energy distribution of the star is expected to be more severe at shorter wavelengths in the wavelength region of L-CH.

Anisotropic defocusing has the potential to provide the benefits of defocusing without not so large loss in spectral resolving power as found with isotropic defocusing by shifting the detectors. To evaluate such anisotropic defocusing, cylindrical aberration is added to the focusing mirrors in simulation to decrease the Strehl ratio at the shortest wavelength of each channel to just 0.2. As a result, additional curvature in the y-direction by $0.0000474\text{mm}^{-1}$ and $0.0001515\text{mm}^{-1}$ is given to Mirror-S and Mirror-L, respectively. Spot diagrams are shown in Fig.5. $R$ and Strehl ratio are summarized in Table A3 in the appendix. It is also possible to give hybrid defocusing by combining simple defocusing with anisotropic defocusing. There are further various defocusing methods, e.g., tilting the detectors, use of additional transimissive optics to make aberration, and so on. Such methods have potential to realize flexible design of wavelength-depended defocusing, however, complexity of the design and calibration should be considered carefully.

## 2.3. Variations in the design

If another transimissive device can be added, the flexibility of the design increases greatly. A design with three prisms, Design #2, is presented in Fig.6. Prism-C has a dichroic coating on its first surface, at which the incident angle of the beam is 30 degrees. It is expected that such a moderate incident angle relaxes the constraints on the design of the dichroic coating. Owing to the dispersion of the two prisms, the separation power of the spectrum order is also improved. The grating pattern is rotated by 11 degree to obtain better perpendicularity between the spectral dispersion and the slit direction for L-CH. Fig. 7 shows the expected spectral dispersion at detector-L without additional defocusing. The effects of defocusing for Designs #2 and #1 were found to be similar, both when the detectors were shifted and with the introduction of cylindrical aberration to the mirrors. The optics for S-CH are common to both Designs #1 and #2.

The designs presented above are just examples, and there are many more variations of the design. First of all, the scientific requirements should be studied well. Simulating observations for a real catalog of targets, the performance of all parts of the observation system including the telescope, the satellite, and especially the detectors, will give the best specifications for the wavelength coverage, the spectral resolving power, the plate scale of the images, defocusing, the operation of the detectors, and so on. Trade-off for the prism material is important, especially for the case of extending the spectral coverage of L-CH toward longer wavelengths. On the other hand, a normal reflective grating is useful if it is possible to decrease the spectral coverage requirement of L-CH by one order. The wavefront of an actual telescope is not perfect, while an ideal point light source is supposed for the design shown above. Fortunately, transiting exoplanets are point-like sources, so the observation does not require a wide field of view. In this situation, it is possible to cancel the expected telescope aberration by design of the mirrors in the spectrometer.

# 3.   Application

JWST and SPICA are large infrared space telescopes with launch dates planned in the near future. Because much of the development of the instruments for JWST has already been executed, SPCIA is a relatively more realistic platform for the wideband infrared spectrometer presented in this paper. Especially, the study of transiting exopalnets is regarded as one of two critical science cases for the SPICA Coronagraph Instrument (SCI), as well as the study of exoplanets with a coronagraph (e.g., Enya et al. 2010; 2011 and their references). The SCI is equipped with a mechanical mask changer including a slot for no mask (i.e., blank hole) for the non-coronagraphic mode. It is planned to carry out observations of transiting exoplanets with the non-coronagraphic mode. Comparing the SCI and another mid-infrared SPICA instrument, the Mid-infrared Camera and Spectrometer (MCS; Kataza et al. 2010), only the SCI is equipped with an InSb detector to cover 1−5 microns. If the SCI adopts an internal tip-tilt mirror (at current, adoption or no adoption of the tip-tilt mirror is TBD), higher pointing accuracy can be realized. These points are potential advantages of the SCI as the platform for a wideband infrared spectrometer for characterizing transiting exoplanets.

If it is eventually decided to have a wideband infrared spectrometer in the SCI, one possible solution of the configuration design is to replace the current short channel of the SCI with the wideband infrared spectrometer, as shown in Fig.8. In this solution, the focal plane in the SCI is shared by the coronagraphic channel and the wideband infrared spectrometer. Channels in this solution are switched by telescope pointing, making it unnecessary to use the mechanism to move the focal plane mask. On the other hand, the fore-optics of the SCI includes a pupil mask changer mechanism. It is considered that the pupil mask changer is equipped with coronagraphic masks (high contrast observation), a blank hole (non-coronagraphic observation), aperture masking (very high spatial resolution), and a stop (calibration). It is planned to use the non-coronagraphic mode for the observation of transiting exoplanets. Another possible function of the fore-optics is the inclusion of a tip-tilt mirror. Though the tip-tilt mirror is a moving part, it is expected to enable more accurate observations of transiting exoplanets by improving the telescope pointing stability. A deformable mirror is another active optical component previously considered to improve the contrast of the coronagraph. The deformable mirror is out of the current baseline design of the SCI because of technical difficulties and the very serious resource limit of the spacecraft, though it has not yet been fully abandoned. Satisfying the constraints of volume and weight is critical for instruments for a space telescope. Unfortunately, the change in design described above increases the number of detectors. On the other hand, it will be possible to remove the filter wheel of the short channel of the current design of the SCI. Reducing the mass and volume of the instrument looks perhaps possible by sophisticated design of the preliminary optics. If the SCI can carry a wideband infrared spectrometer, it might be reasonable to change the name of the instrument to express its "high dynamic range", rather than its "coronagraphic" properties. It would also be possible, in principle, to realize a wideband infrared spectrometer as a part of MCS, as an independent instrument.

JWST is equipped with the Near-Infrared Spectrograph (NIRSPEC) for wavelengths less than 5 microns and the Mid-Infrared Instrument (MIRI) for wavelengths greater than 5 microns (e.g., Clampin 2008). Spectroscopic observations and monitoring of transiting exoplanets is possible with both NIRSPEC and MIRI. JWST has a larger telescope aperture (6.5m) than SPICA (3m class). On the other hand, simultaneous use of NIRCAM and MIRI for one point-like target is impossible, and simultaneous wavelength coverage by each instrument is limited. In principle, defocusing is possible by adding a diffuser into the instruments, moving the secondary mirror of the telescope, and/or actuation of the segmented primary mirror, though it is usually hard to alter the optical alignment of a space telescope mirror for a peculiar observation.

It should be noted that proposed future missions: The terrestrial habitable-zone exoplanet spectroscopy infrared spacecraft (e.g., THESIS; Swain et al. 2010), with 1.4m telescope and spectrometer was proposed for the observation of transiting exoplanets (see also the Fast Infrared

Exoplanet Spectroscopy Survey Explorer, FINESSE; e.g., Swain 2010). The Exoplanet Characterization Observatory (EChO), is optimized mission for the study of transiting exoplanets. EChO adopts a 1.5m class telescope with spectrometers, and covers the wavelength region of 0.4−16 microns (e.g., Tinetti et al. 2012). If SPICA will the wideband infrared spectrometer, it will be complementally with these dedicated missions because of difference of stability of the system (available stability/dedicated stability), telescope time (part time of multi-purpose mission/full time) , and the size of the telescope aperture.

## 4.   Final remark

Based on the concept and discussion given in this paper, the following two points are proposed for SPICA: 1) to develop and carry a wideband infrared spectrometer optimized for the observation of transiting exoplanets, 2) to perform dedicated science on transiting exoplanets. To realize this scheme, it is considered there are many issues to be completed: the science goals should be clearly defined. The specification of the instrument and the error budget should be derived from the science requirements, and should be optimized by analyzing the performance of all parts of the observation system, including the telescope and the satellite by simulating observations of a real catalog of targets. Especially, knowledge and simulation of the performance of the detectors, including the stability, homogeneity, and influence of hitting of cosmic ray, are important. Both simulation and laboratory demonstration are necessary. From a practical point of view, the design of the instrument needs to be compact and lightweight in order to satisfy the severe constraints of the resources in the spacecraft. Additional folding mirrors are useful.

It should be noted that the possibility of detecting biomarkers, including ozone and/or oxygen, on exoplanets with the SCI is not zero, though the current primary target of the SCI is Jovian exoplanets and the detection of biomarkers is not robustly expected. To enable such detection with the total functions of the SCI, capability to get the spectrum (or spectral energy distribution, at least) is important. The coronagraphic mode has the advantage of high contrast observations, the aperture masking method has potential to provide higher spatial resolution, and transit monitoring can target exoplanets which are not spatially resolved from the central star. These modes are complementally. Spectroscopic functions working with each of these modes should be realized. For the spatially resolved observations, targeting very nearby stars and/or early type stars is interesting in order to approach the habitable zone (HZ), because the former has an HZ with apparently large angular size and the latter has an absolutely large HZ. The possibility of finding biomarkers even in the classical "HZ" should not be ignored, e.g., in the atmospheres of outer exoplanets, eccentrically orbiting exoplanets, exo-satellites, and so on. The pre-discovery of targets is also essential for subsequently characterizing transiting exoplanets using dedicated space infrared telescopes.

For future space missions designed to monitor transiting exoplanets, there is the potential to re-use the resources from previous or ongoing missions. Especially useful resources are the mirrors that have already been manufactured, e.g., the backup mirrors for HERSCHEL (3.5m: e.g., ref. [11]), SPICA (3m class), AKARI (68.5cm: e.g., Murakami et al.2006), or mirrors manufactured for technical demonstrations. Usually, polishing the mirrors takes a large fraction of the time needed for mission development. However, special missions to monitor transiting exoplanet do not require mirrors with the highest quality surfaces. Other important resources are the systems of such satellites. There is also the opportunity to adopt significantly common bus-systems in a series of missions, e.g., in the small satellite program of JAXA. Such an approach would reduce cost, delivery time, and risk, and increase the feasibility and possibility of realizing missions.

# References

[1] Clampin, M., 2008, proc. of SPIE, 7010, 0L

[2 ] Enya, K. SPICA Working Group, 2010, AdSpR, 45, 979

[3] Enya, K., Abe, L., Takeuchi, S., Kotani, T., Yamamuro, T., 2011, proc. of SPIE, 8146, 0Q

[4] Enya, K., Yamada, N., Onaka, T., Nakagawa, T., Kaneda, H., Hirabayashi, M., Toulemont, Y., Castel, D., Kanai, Y., Fujishiro, N., 2007, PASP, 119, 583

[5] Enya, K., Yamada, N., Imai, T., Tange, Y., Kaneda, H., Katayama, H., Kotani, M., Maruyama, K., Naitoh, M., Nakagawa, T., Onaka, T., Suganuma, M., Ozaki, T., Kume, M., Krödel, M. R., 2012, Cryogenics, 52, 86

[6] Kataza, H., Wada, T., Ikeda, Y., Fujishiro, N., Kobayashi, N., Sakon, I., 2010, proc. of SPIE, 7731, 4A

[7] Nakagawa, T., 2010, proc. of SPIE, 7731, 0O

[8] Swain, M. R., Vasisht, G., Henning, T., Tinetti, G., Beaulieu, J. P., proc. of SPIE, 2010, 7731, 773125

[9] Swain, M. R., 2010, American Astronomical Society, DPS meeting #42, #27.22; Bulletin of the American Astronomical Society, Vol. 42, p.1064

[10] Tinetti, G., et al., 2011, astro-ph, arXiv:1112.2728

[11] http://www.esa.int/SPECIALS/Herschel/index.html

[12] Murakami, H., et al., 2007, 59, 369

# Tables

Table 1. Summary for Design #1.

| | | |
|---|---|---|
| Collimator optics | Beam diameter | 6mm |
| | Focal length | 69mm |
| S-CH | Prism-S material | ZnSe |
| | Prism-S apex angle | 40 degree |
| | Grating pitch | 55 micron |
| | Focal length | 158mm |
| | Detector pixel size | 25um × 25um |
| | Detector format | 1k × 1k |
| L-CH | Prism-L material | ZnS |
| | Prism-L apex angle | 45 degree |
| | Grating pitch | 77.5 micron |
| | Focal length | 96mm |
| | Detector pixel size | 30um × 30um |
| | Detector format | 1k × 1k |

Table 2. Summary for Design #2.

| | | |
|---|---|---|
| Collimator optics | Beam diameter | 6mm |
| | Focal length | 69mm |
| S-CH | Prism-S material | ZnSe |
| | Prism-S apex angle | 40 degree |
| | Grating pitch | 55 micron |
| | Focal length | 158mm |
| | Detector pixel size | 25um × 25um |
| | Detector format | 1k × 1k |
| L-CH | Prism-L1 material | ZnS |
| | Prism-L1 apex angle | 20 degree |
| | Prism-L2 material | ZnS |
| | Prism-L2 apex angle | 45 degree |
| | Grating pitch | 65 micron |
| | Focal length | 96mm |
| | Detector pixel size | 30um × 30um |
| | Detector format | 1k × 1k |

# Figures

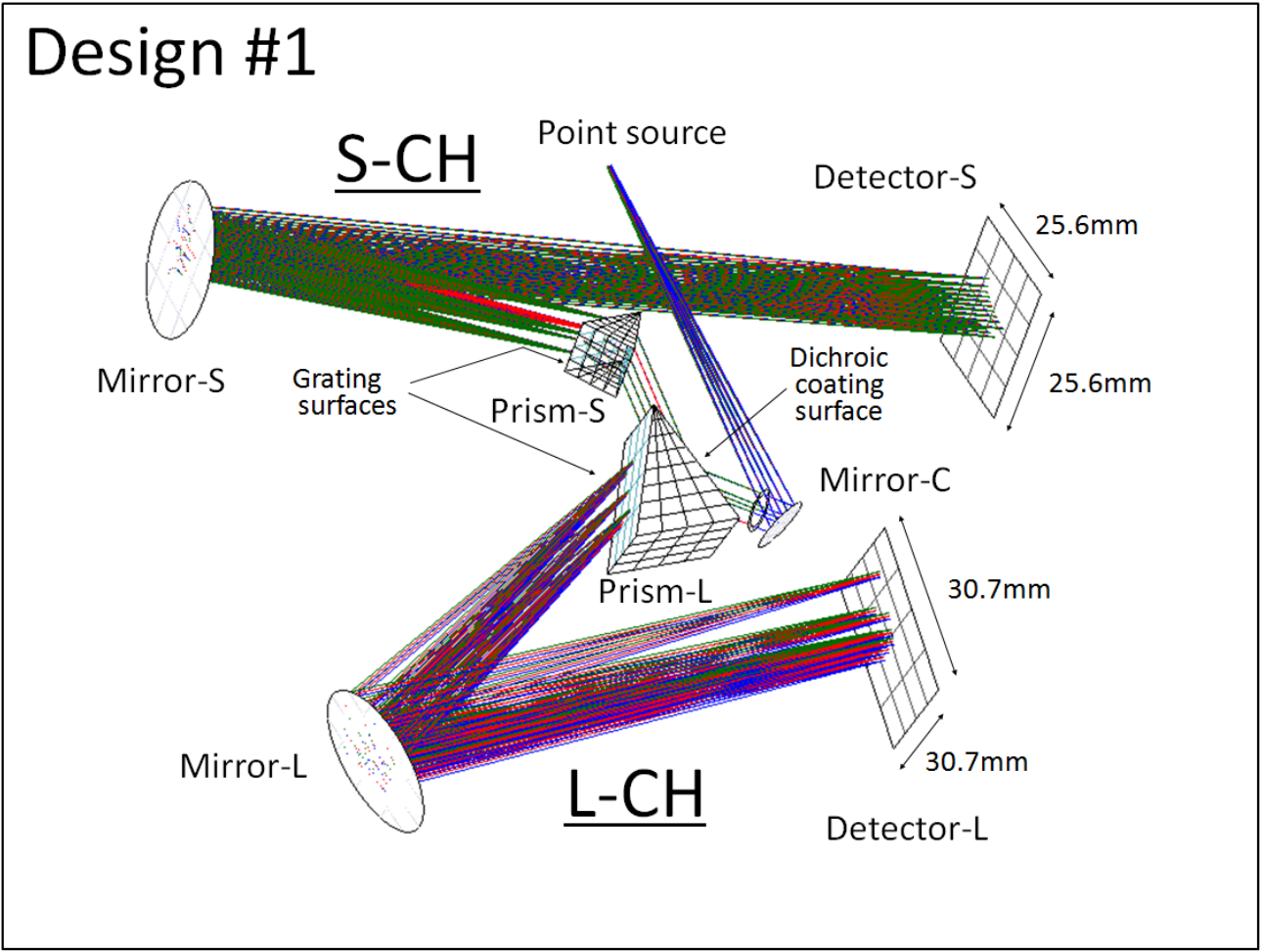

Fig1. Overview of Design #1. A ring after Mirror-C is imaginary one introduced to define the pupil of this optics.

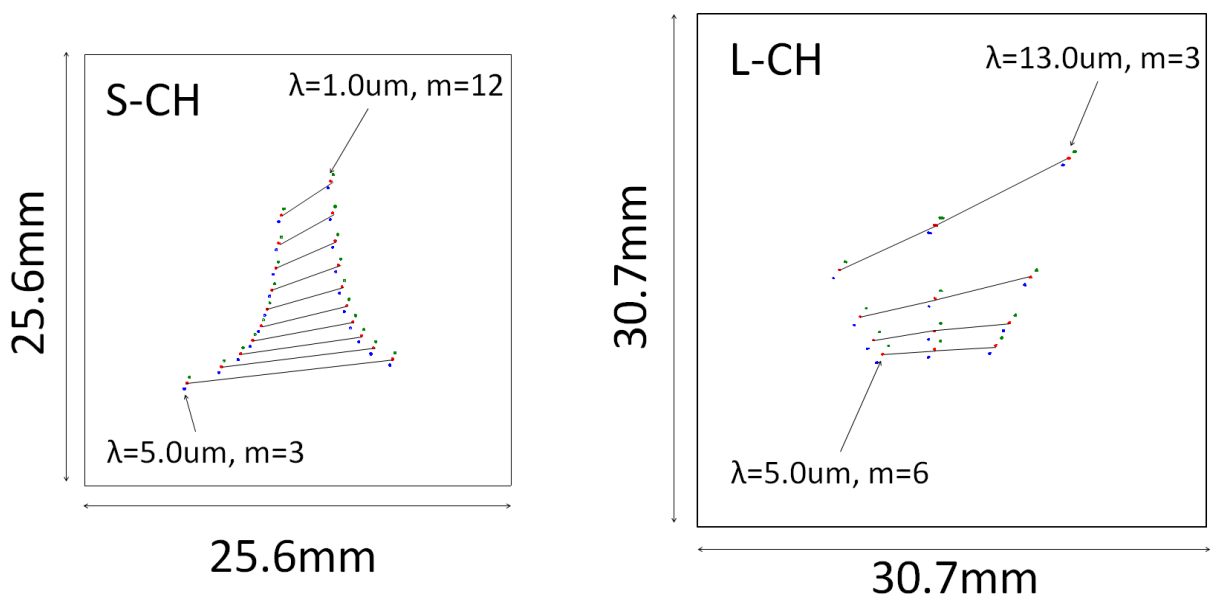

Fig.2. Spectral dispersion obtained by simulation using Design #1. The green and red plots show the slit length at the focal plane corresponding to +/- 2 arcsec in the direction of the prism dispersion.

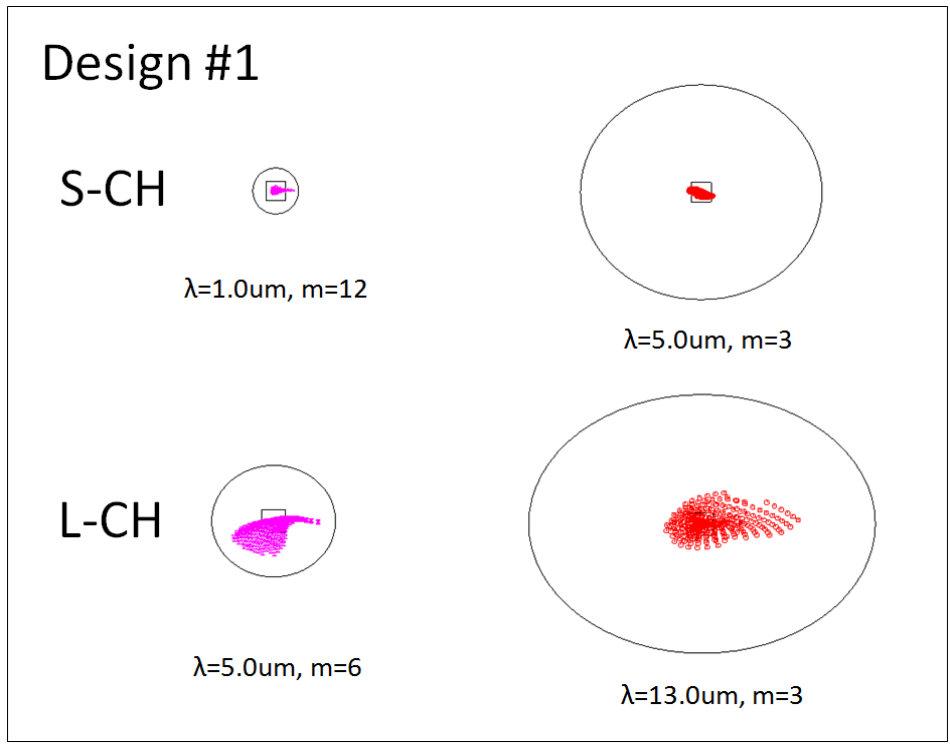

Fig.3 Spot diagrams obtained by simulation using Design #1. Ellipses and squares show Airy disk and pixel size, respectively.

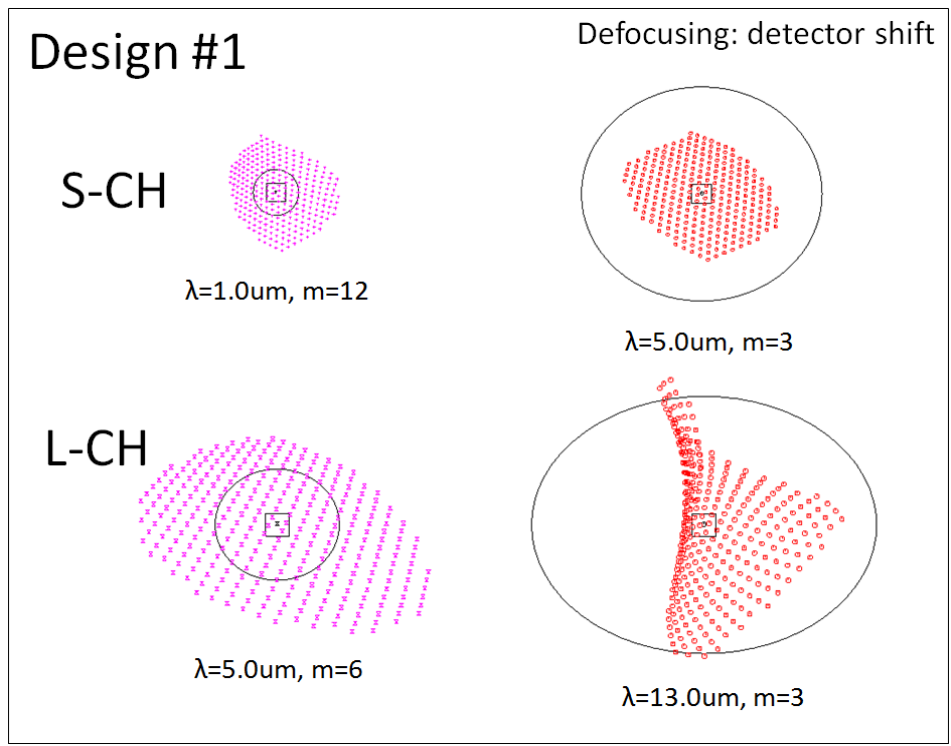

Fig 4. Spot diagrams obtained by simulation based on Design #1 but with defocusing by shifting the detector. Ellipses and squares show Airy disk and pixel size, respectively.

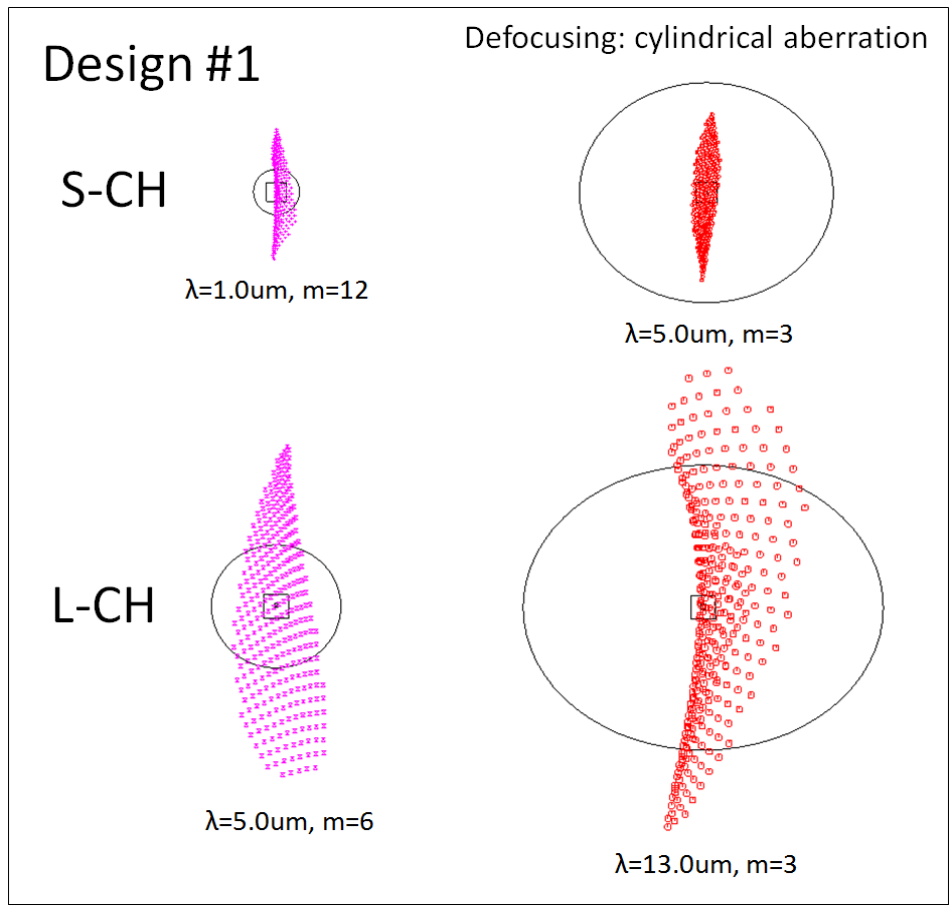

Fig 5. Spot diagrams obtained by simulation based on Design #1 but with defocusing by cylindrical aberration of the mirror. Ellipses and squares show Airy disk and pixel size, respectively.

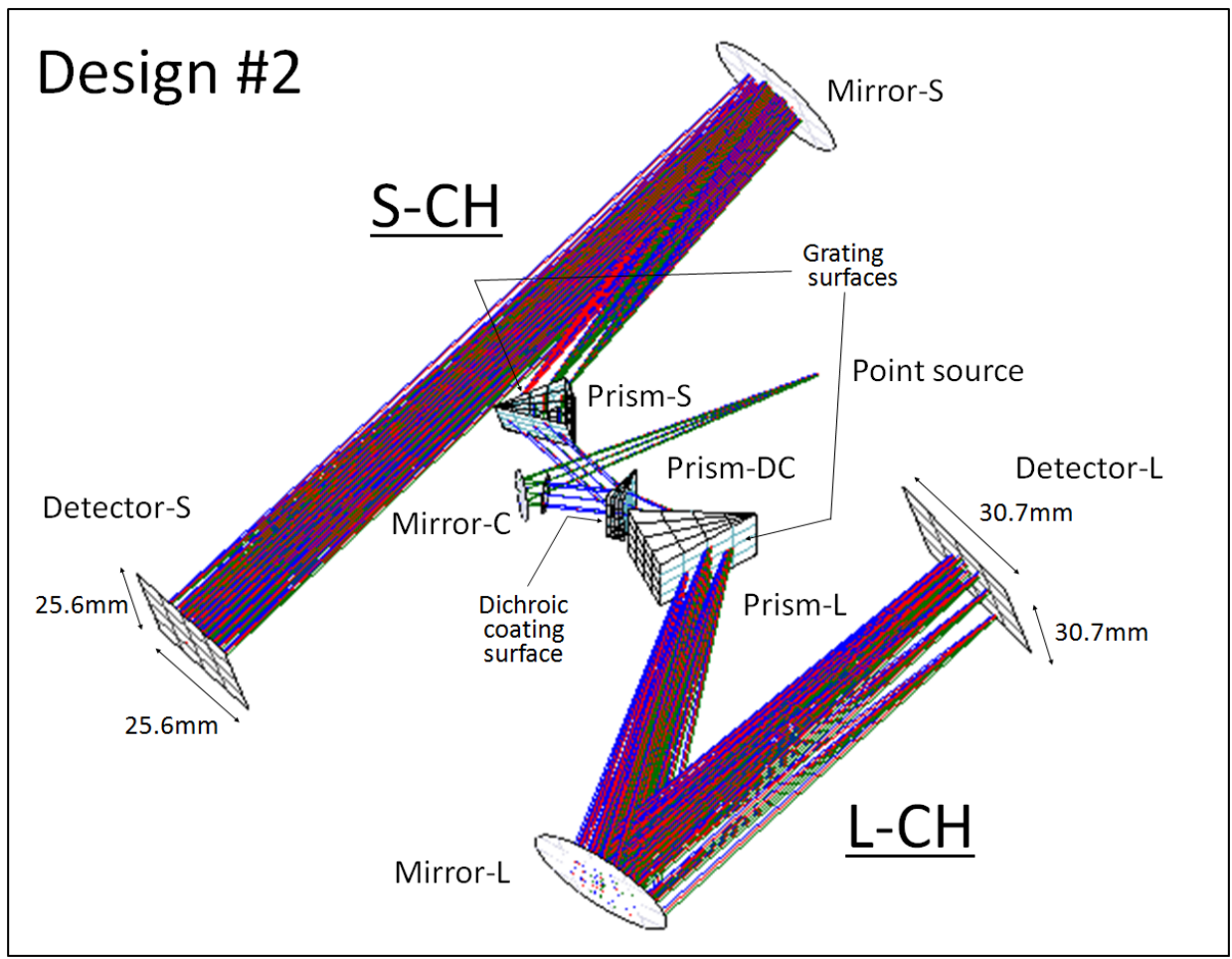

Fig. 6 Overview of Design #2. A ring after Mirror-C is imaginary one introduced to define the pupil of this optics.

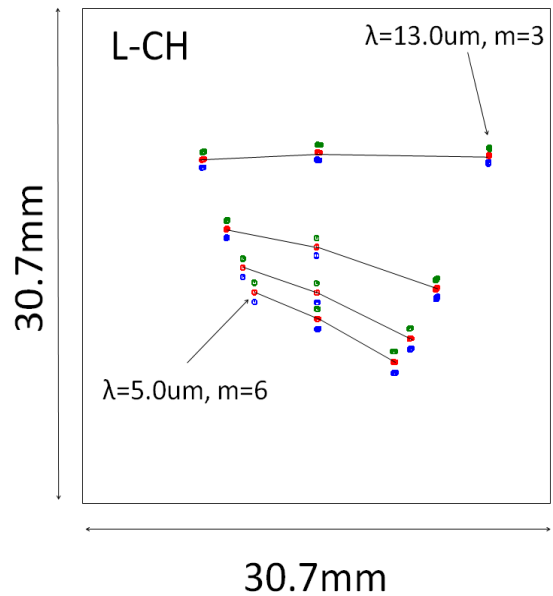

Fig. 7 Spectral dispersion of the L-CH obtained by simulation using Design #2. The geen and red plots show the slit length at the focal plane corresponding to +/- 2 arcsec in the direction of the prism dispersion.

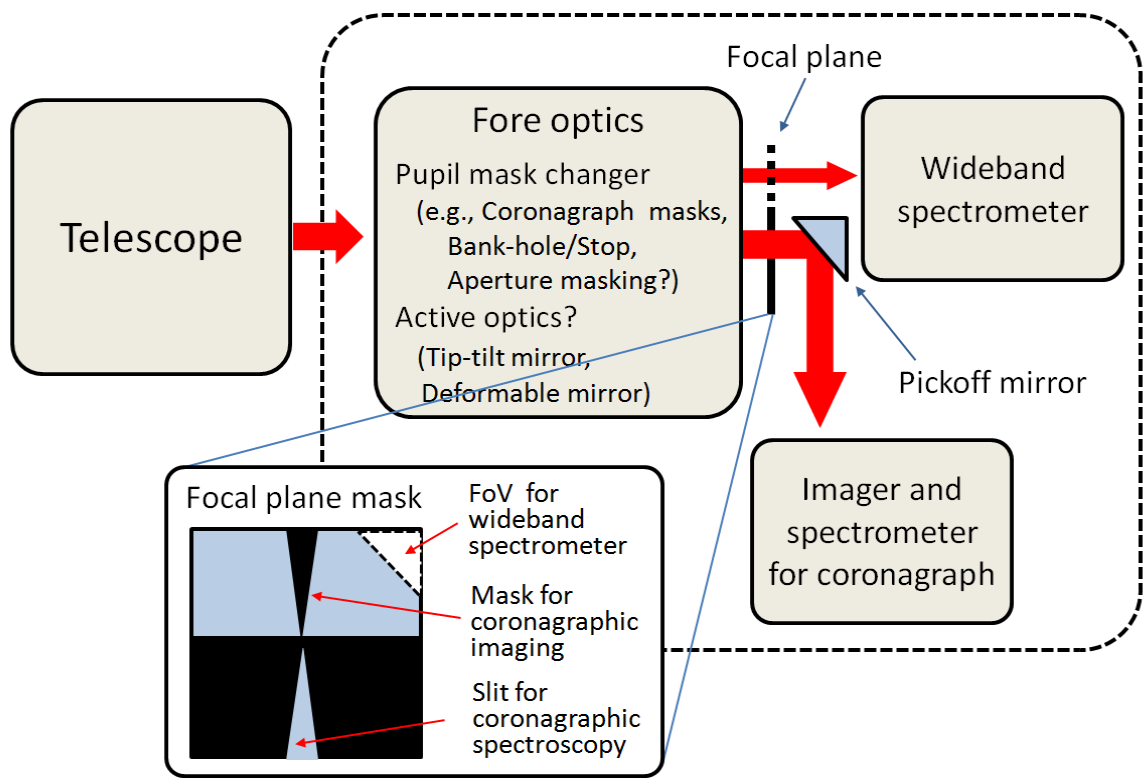

Fig. 8 Schematic view of a possible configuration in which a wideband spectrometer is added to the SPICA-SCI. The channel is switched on the focal plane mask by telescope pointing without the need for a mechanism to move the mask. The detailed design of the mask and slits are TBD.

# Appendix

Table A1. Spectral resolving power with Design #1.

| Chanel | Order of spectrum | Wavelength (nm) | R | | | Strehl Ratio |
|---|---|---|---|---|---|---|
| | | | 1.22× lambda ×F | 2×(Pixel Size) | RMS SPOT | |
| S-CH | 3 | 5000 | **311.377** | 864.77 | 2670.32 | 0.996783 |
| | 3 | 3571 | **330.453** | 628.013 | 2057.86 | 0.99614 |
| | 4 | 3570 | **418.044** | 823.552 | 3499.07 | 0.99714 |
| | 4 | 2778 | **438.625** | 651.562 | 2464.7 | 0.994867 |
| | 5 | 2777 | **525.306** | 802.127 | 4069.77 | 0.997026 |
| | 5 | 2273 | **548.96** | 669.322 | 2932.33 | 0.994362 |
| | 6 | 2271 | **633.94** | 790.46 | 4833.12 | 0.997282 |
| | 6 | 1923 | **662.693** | 685.315 | 3557.38 | 0.994576 |
| | 7 | 1922 | **745.209** | 785.359 | 6115.55 | 0.998093 |
| | 7 | 1666 | 781.859 | **702.142** | 4307.88 | 0.995021 |
| | 8 | 1665 | 861.065 | **785.988** | 7943.78 | 0.998836 |
| | 8 | 1470 | 909.464 | **721.977** | 4918.98 | 0.995035 |
| | 9 | 1469 | 984.397 | **792.681** | 7949.04 | 0.998185 |
| | 9 | 1315 | 1049.76 | **747.004** | 5000.32 | 0.993521 |
| | 10 | 1314 | 1119.35 | **806.561** | 5814.5 | 0.993847 |
| | 10 | 1190 | 1208.57 | **779.646** | 4565.32 | 0.989644 |
| | 11 | 1189 | 1271.72 | **829.424** | 4076.7 | 0.982407 |
| | 11 | 1086 | 1393.69 | **822.738** | 4007.06 | 0.982935 |
| | 12 | 1085 | 1449.4 | **863.749** | 3006.82 | 0.959178 |
| | 12 | 1000 | 1615.38 | **879.686** | 3445.61 | 0.971 |
| L-CH | 3 | 13000 | **312.234** | 1098.4 | 742.725 | 0.979776 |
| | 3 | 10833 | **318.5** | 883.067 | 355.195 | 0.869952 |
| | 3 | 9285 | **325.193** | 760.497 | 1006.44 | 0.985439 |
| | 4 | 9284 | **373.258** | 944.112 | 619.227 | 0.928796 |
| | 4 | 8125 | **386.866** | 824.717 | 772.451 | 0.955037 |
| | 4 | 7222 | **399.08** | 740.832 | 1285.83 | 0.992244 |
| | 5 | 7221 | **456.619** | 902.656 | 658.633 | 0.915591 |
| | 5 | 6500 | **471.121** | 812.484 | 901.281 | 0.960546 |
| | 5 | 5909 | **484.114** | 743.899 | 857.939 | 0.969399 |
| | 6 | 5908 | **545.407** | 883.407 | 636.38 | 0.883789 |
| | 6 | 5416 | **559.73** | 809.341 | 756.254 | 0.931647 |
| | 6 | 5000 | **572.782** | 750.433 | 676.872 | 0.924527 |

Table A2. Spectral resolving power based on Design #1 but with defocusing by shifting the detectors.

| Chanel | Order of spectrum | Wavelength (nm) | R | | | Strehl Ratio |
| --- | --- | --- | --- | --- | --- | --- |
| | | | 1.22× lambda ×F | 2×(Pixel Size) | RMS SPOT | |
| S-CH (defocusing by 4.00mm detector shift) | 3 | 5000 | **307.451** | 853.867 | 349.211 | 0.901468 |
| | 3 | 3571 | 326.287 | 620.096 | **272.837** | 0.843146 |
| | 4 | 3570 | 412.771 | 813.164 | **337.897** | 0.827885 |
| | 4 | 2777 | 433.091 | 643.341 | **280.07** | 0.750946 |
| | 5 | 2776 | 518.678 | 792.006 | **332.598** | 0.734224 |
| | 5 | 2272 | 542.03 | 660.873 | **287.903** | 0.650091 |
| | 6 | 2271 | 625.939 | 780.483 | **331.385** | 0.642129 |
| | 6 | 1923 | 654.325 | 676.661 | **296.732** | 0.548964 |
| | 7 | 1922 | 735.801 | 775.444 | **333.639** | 0.549346 |
| | 7 | 1666 | 771.982 | 693.272 | **307.344** | 0.460261 |
| | 8 | 1665 | 850.191 | 776.063 | **339.473** | 0.463948 |
| | 8 | 1470 | 897.971 | 712.854 | **320.749** | 0.391795 |
| | 9 | 1469 | 971.964 | 782.67 | **349.534** | 0.400679 |
| | 9 | 1315 | 1036.49 | 737.561 | **338.182** | 0.332262 |
| | 10 | 1314 | 1105.21 | 796.372 | **364.955** | 0.334028 |
| | 10 | 1190 | 1193.29 | 769.788 | **359.486** | 0.272906 |
| | 11 | 1189 | 1255.66 | 818.945 | **387.331** | 0.278979 |
| | 11 | 1086 | 1376.06 | 812.332 | **389.157** | 0.228467 |
| | 12 | 1085 | 1431.09 | 852.835 | **418.577** | 0.251794 |
| | 12 | 1000 | 1594.95 | 868.559 | **426.314** | 0.202 |
| L-CH (defocusing by 3.55mm detector shift) | 3 | 13000 | 309.113 | 1087.41 | **278.594** | 0.812444 |
| | 3 | 10833 | 315.44 | 874.584 | **239.268** | 0.757391 |
| | 3 | 9285 | 322.057 | 753.162 | **184.93** | 0.666787 |
| | 4 | 9284 | 369.666 | 935.026 | **236.275** | 0.767019 |
| | 4 | 8125 | 383.2 | 816.901 | **212.179** | 0.701534 |
| | 4 | 7222 | 395.292 | 733.799 | **168.392** | 0.481985 |
| | 5 | 7221 | 452.307 | 894.132 | **215.578** | 0.626482 |
| | 5 | 6500 | 466.688 | 804.838 | **196.223** | 0.530446 |
| | 5 | 5909 | 479.549 | 736.884 | **164.219** | 0.327485 |
| | 6 | 5908 | 540.299 | 875.133 | **204.75** | 0.466822 |
| | 6 | 5416 | 554.485 | 801.757 | **188.122** | 0.361105 |
| | 6 | 5000 | 567.401 | 743.383 | **162.761** | 0.202273 |

Table A3. Spectral resolving power based on Design-1 but with defocusing by cylindrical aberration of the mirrors.

| Chanel | Order of spectrum | wavlength (nm) | R | | | | | Strehl Ratio |
| --- | --- | --- | --- | --- | --- | --- | --- | --- |
| | | | 1.22× Lambda ×F | 2×(Pixel Size) | RMS SPOT | | | |
| | | | | | (XY) | X | (Y) | |
| S-CH (Y-direction curvature of focusing mirror +0.0000474/mm) | 3 | 5000 | **312.948** | 864.71 | 433.856 | 2914.53 | 438.744 | 0.882492 |
| | 3 | 3571 | **331.781** | 627.696 | 317.994 | 2663.78 | 320.285 | 0.817131 |
| | 4 | 3570 | **420.082** | 823.484 | 423.137 | 3626.75 | 426.047 | 0.797704 |
| | 4 | 2777 | **440.383** | 651.224 | 333.944 | 3195.92 | 335.783 | 0.720542 |
| | 5 | 2776 | **527.782** | 802.016 | 418.245 | 4198.21 | 420.336 | 0.719762 |
| | 5 | 2272 | **551.073** | 668.887 | 349.198 | 3799.1 | 350.683 | 0.624038 |
| | 6 | 2271 | **636.8** | 790.263 | 418.491 | 5041.49 | 419.941 | 0.617213 |
| | 6 | 1923 | **665.053** | 684.703 | 364.721 | 4709.7 | 365.82 | 0.53243 |
| | 7 | 1922 | **748.361** | 785.013 | 422.863 | 6611.77 | 423.73 | 0.518634 |
| | 7 | 1666 | 784.305 | **701.254** | 381.148 | 5950.32 | 381.933 | 0.449111 |
| | 8 | 1665 | 864.364 | **785.403** | 430.698 | 9429.1 | 431.148 | 0.465722 |
| | 8 | 1470 | 911.759 | **720.683** | 398.961 | 6871.66 | 399.636 | 0.373329 |
| | 9 | 1469 | 987.624 | **791.735** | 441.621 | 10168.7 | 442.038 | 0.394847 |
| | 9 | 1315 | 1051.57 | **745.135** | 418.691 | 6392.77 | 419.592 | 0.321167 |
| | 10 | 1314 | 1122.19 | **805.087** | 455.533 | 6590.56 | 456.625 | 0.355922 |
| | 10 | 1190 | 1209.43 | **776.991** | 441.25 | 5211.14 | 442.841 | 0.254807 |
| | 11 | 1189 | 1273.72 | **827.211** | 472.835 | 4282.95 | 475.743 | 0.312748 |
| | 11 | 1086 | 1393 | **819.037** | 468.542 | 4230.2 | 471.442 | 0.253654 |
| | 12 | 1085 | 1449.94 | **860.53** | 494.96 | 3089.46 | 501.437 | 0.254195 |
| | 12 | 1000 | 1612.35 | **874.629** | 504.592 | 3531.51 | 509.823 | 0.200316 |
| L-CH (Y-direction Curvature of focusing mirror +0.0001515/mm) | 3 | 13000 | **306.605** | 1108.15 | 196.473 | 796.703 | 202.734 | 0.6189 |
| | 3 | 10833 | **313.913** | 890.152 | 145.426 | 367.936 | 158.317 | 0.517533 |
| | 3 | 9285 | **320.99** | 766.398 | 146.194 | 1304.07 | 147.121 | 0.520765 |
| | 4 | 9284 | **365.087** | 945.537 | 181.179 | 790.769 | 186.13 | 0.511295 |
| | 4 | 8125 | **379.472** | 826.526 | 162.072 | 916.4 | 164.667 | 0.476759 |
| | 4 | 7222 | **392.21** | 742.962 | 150.673 | 1657.68 | 151.299 | 0.391007 |
| | 5 | 7221 | **446.653** | 902.924 | 185.673 | 895.549 | 189.797 | 0.514257 |
| | 5 | 6500 | **461.691** | 813.093 | 172.339 | 1195.88 | 174.157 | 0.400327 |
| | 5 | 5909 | **475.131** | 744.857 | 158.475 | 963.983 | 160.661 | 0.292263 |
| | 6 | 5908 | **533.753** | 883.501 | 191.231 | 838.233 | 196.411 | 0.398834 |
| | 6 | 5416 | **548.463** | 809.614 | 179.364 | 948.075 | 182.663 | 0.309205 |
| | 6 | 5000 | **561.884** | 750.954 | 165.146 | 749.645 | 169.305 | 0.203588 |

Table A4. Spectral resolving power with Design #2.

| Chanel | Order of spectrum | wavelength (nm) | R | | | Strehl Ratio |
| --- | --- | --- | --- | --- | --- | --- |
| | | | 1.22× Lambda ×F | 2*(Pixel Size) | RMS SPOT | |
| L-CH | 3 | 13000 | **356.557** | 1369.99 | 1274.83 | 0.981402 |
| | 3 | 10833 | **295.789** | 1088.07 | 464.842 | 0.909376 |
| | 3 | 9285 | **272.291** | 890.331 | 491.896 | 0.911403 |
| | 4 | 9285 | **489.573** | 1407.09 | 625.835 | 0.848884 |
| | 4 | 8125 | **412.087** | 1106.87 | 1893.21 | 0.991162 |
| | 4 | 7222 | **369.939** | 932.869 | 799.782 | 0.946051 |
| | 5 | 7222 | **622.793** | 1436.39 | 960.578 | 0.900241 |
| | 5 | 6500 | **533.994** | 1154.51 | 1327.1 | 0.96065 |
| | 5 | 5909 | **489.363** | 986.76 | 1306.26 | 0.972944 |
| | 6 | 5909 | **757.788** | 1439.88 | 791.693 | 0.828466 |
| | 6 | 5416 | **663.19** | 1191.63 | 816.52 | 0.864704 |
| | 6 | 5000 | **603.591** | 1032.3 | 1721.69 | 0.985946 |

Table A6. Spectral resolving power based on Design #2 but with defocusing by shifting the detectors.

| Chanel | Order of spectrum | wavelength (nm) | R | | | Strehl Ratio |
| --- | --- | --- | --- | --- | --- | --- |
| | | | 1.22× Lambda ×F | 2*(Pixel Size) | RMS SPOT | |
| L-CH (defocusing by 5.75mm detector shift) | 3 | 13000 | 333.652 | 1354.38 | **233.596** | 0.604775 |
| | 3 | 10833 | 281.387 | 1073.32 | **148.001** | 0.39518 |
| | 3 | 9285 | 260.969 | 878.302 | **127.658** | 0.350161 |
| | 4 | 9285 | 482.034 | 1385.42 | **341.169** | 0.696508 |
| | 4 | 8125 | 406 | 1090.52 | **212.039** | 0.416422 |
| | 4 | 7222 | 342.392 | 919.314 | **154.276** | 0.310974 |
| | 5 | 7222 | 632.991 | 1413.57 | **341.589** | 0.470989 |
| | 5 | 6500 | 552.754 | 1136.86 | **248.046** | 0.33733 |
| | 5 | 5909 | 585.647 | 971.934 | **178.183** | 0.200236 |
| | 6 | 5909 | 764.679 | 1417.05 | **355.129** | 0.368132 |
| | 6 | 5416 | 672.817 | 1173.07 | **272.109** | 0.29705 |
| | 6 | 5000 | 619.357 | 1016.46 | **200.383** | 0.222056 |

Table A7. Spectral resolving power based on Design 2 but with defocusing by cylindrical aberration of the mirrors.

| Chanel | Order of spectrum | wavlength (nm) | R | | | | | Strehl Ratio |
| --- | --- | --- | --- | --- | --- | --- | --- | --- |
| | | | 1.22× Lambda ×F | 2*(Pixel Size) | RMS SPOT | | | |
| | | | | | (×Y) | X | (Y) | |
| L-CH (Y-direction Curvature of focusing mirror +0.00044/mm) | 3 | 13000 | **355.836** | 1371.72 | 139.056 | 2515.89 | 139.269 | 0.565897 |
| | 3 | 10833 | **309.084** | 1089.54 | 108.97 | 463.947 | 112.107 | 0.275296 |
| | 3 | 9285 | **279.7** | 892.354 | 84.8698 | 525.612 | 85.9983 | 0.250031 |
| | 4 | 9285 | **494.552** | 1431.46 | 182.173 | 1010.75 | 185.206 | 0.654627 |
| | 4 | 8125 | **415.357** | 1115.22 | 127.362 | 5664.74 | 127.394 | 0.340736 |
| | 4 | 7222 | **376.633** | 935.694 | 97.2137 | 1084.25 | 97.6068 | 0.250936 |
| | 5 | 7222 | **642.373** | 1477.54 | 180.654 | 1055.16 | 183.362 | 0.473023 |
| | 5 | 6500 | **545.899** | 1173.8 | 132.734 | 1306.02 | 133.425 | 0.277226 |
| | 5 | 5909 | **491.643** | 995.946 | 108.088 | 2315.12 | 108.206 | 0.228438 |
| | 6 | 5909 | **784.972** | 1489.47 | 176.466 | 790.581 | 181.034 | 0.328335 |
| | 6 | 5416 | **680.339** | 1218.76 | 136.321 | 820.738 | 138.241 | 0.251574 |
| | 6 | 5000 | **613.619** | 1047.34 | 116.048 | 2384.33 | 116.185 | 0.20248 |